\documentclass{PoS}
\pdfoutput=1

\title{Search for $K^+\to\pi^+\nu\bar{\nu}$ at NA62}

\ShortTitle{Search for $K^+\to\pi^+\nu\bar{\nu}$ at NA62}

\author{\speaker{Matthew Moulson} for the NA62 Collaboration%
  \thanks{
G.~Aglieri Rinella, R.~Aliberti, F.~Ambrosino, R.~Ammendola, B.~Angelucci, 
A.~Antonelli, G.~Anzivino, R.~Arcidiacono, I.~Azhinenko, 
S.~Balev
% ...
  }\\
        INFN Laboratori Nazionali di Frascati, Italy\\
        E-mail: \email{moulson@lnf.infn.it}}

\abstract{
The decay $K^+\to\pi^+\nu\bar{\nu}$ is highly suppressed in the Standard Model
(SM), while its rate can be predicted with minimal theoretical uncertainty.
The branching ratio (BR) for this decay is thus a sensitive probe of the
flavor sector of the SM; its measurement, however, is a significant
experimental challenge. The primary goal of the NA62 experiment is to measure
${\rm BR}(K^+\to\pi^+\nu\bar{\nu})$ with 10\% precision.
NA62 took data in pilot runs in 2014 and 2015, reaching the design
beam intensity. The status of the experiment and the prospects for the
measurement of ${\rm BR}(K^+\to\pi^+\nu\bar{\nu})$ are presented.
}

\FullConference{38th International Conference on High Energy Physics\\
		3-10 August 2016\\
		Chicago, USA}

\begin{document}

\section{Introduction}

\begin{figure}[ht]
\centering
\includegraphics[width=80mm]{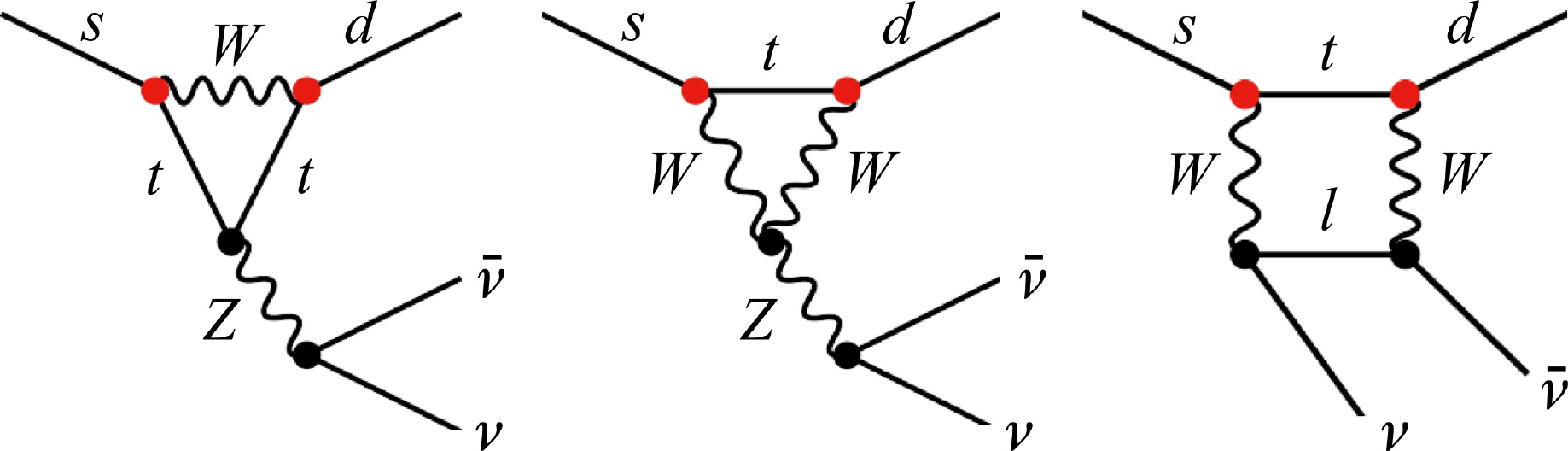}
\caption{Diagrams contributing to the process $K\to\pi\nu\bar{\nu}$.} 
\label{fig:fcnc}
\end{figure}

The $K\to\pi\nu\bar{\nu}$ decays are flavor-changing neutral current (FCNC)
processes that probe the $s\to d\nu\bar{\nu}$ transition via the 
$Z$-penguin and box diagrams shown in Figure~\ref{fig:fcnc}. They are 
highly GIM suppressed and their Standard Model (SM) rates are very small.
For several reasons, the SM calculation for their branching ratios
(BRs) is particularly clean: the loop amplitudes are dominated by the
top-quark contributions, the hadronic matrix element for these decays
can be obtained from the precise experimental measurement of the $K_{e3}$
rate, and there are no long-distance contributions from processes with
intermediate photons (see \cite{Cirigliano:2011ny} for a review).
In the SM,
${\rm BR}(K^+\to\pi^+\nu\bar{\nu}) = (8.4 \pm 1.0)\times10^{-11}$ and
${\rm BR}(K_L\to\pi^0\nu\bar{\nu}) = (3.4 \pm 0.6)\times10^{-11}$
\cite{Buras:2015qea}.
The uncertainties are entirely dominated by the CKM inputs, which in
this case are from tree-level observables. Without these parametric errors,
the uncertainties would be just $0.30\times10^{-11}$ (3.5\%)
and $0.05\times10^{-11}$ (1.5\%), respectively. Because of corrections to
the amplitudes from loops with the lighter quarks, the intrinsic uncertainty
is slightly larger for the charged channel. 

\begin{figure}[ht]
\centering
\includegraphics[width=80mm]{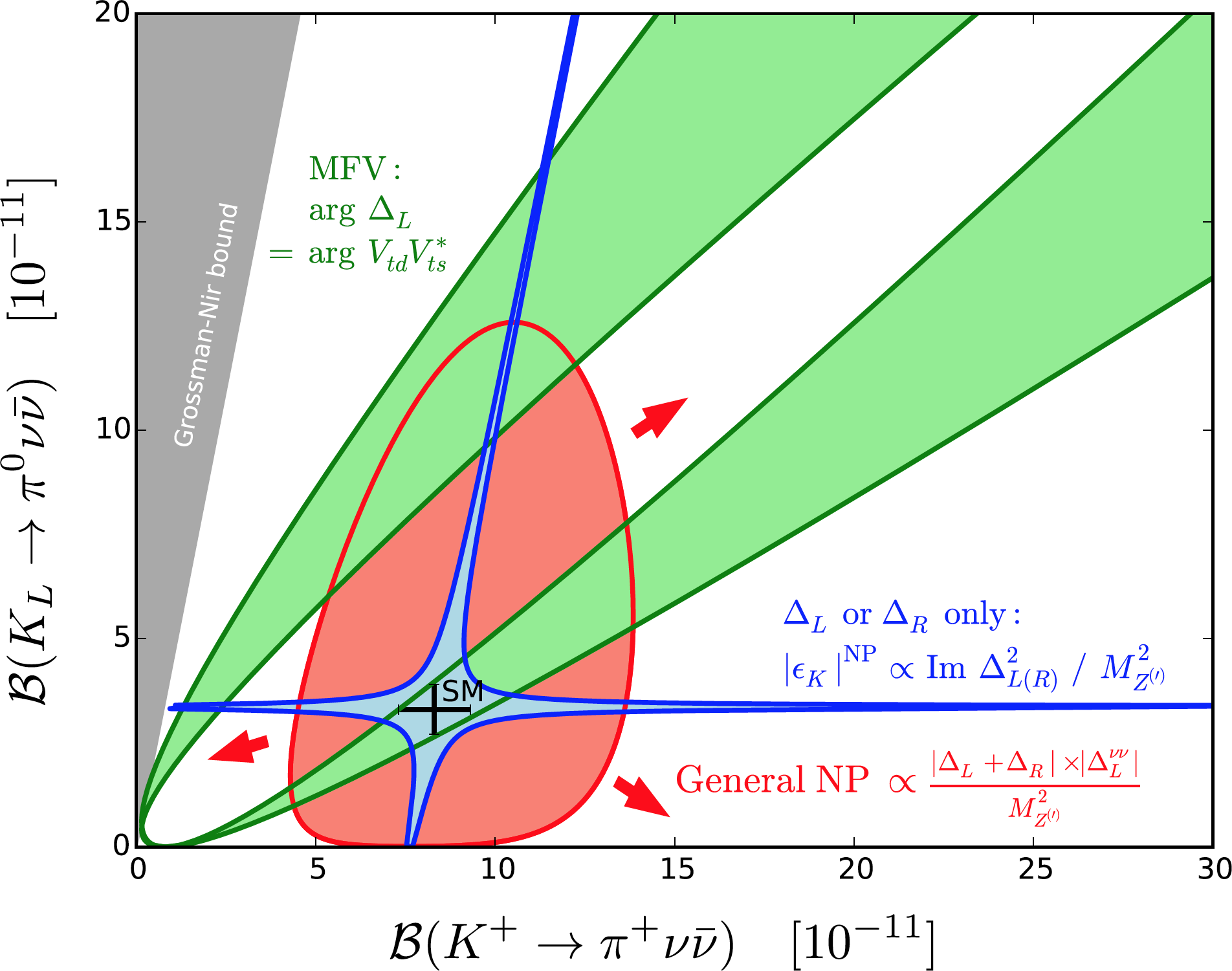}
\caption{Schematic illustration of correlations between BRs for
  $K^+\to\pi^+\nu\bar{\nu}$ and $K_L\to\pi^0\nu\bar{\nu}$ expected
  in different new-physics scenarios, from \cite{Buras:2015yca}.}
\label{fig:bsm}
\end{figure}
Because the SM rates are small and predicted very precisely,
the BRs for these decays are sensitive probes for new physics. 
In general, ${\rm BR}(K_L\to\pi^0\nu\bar{\nu})$ and
${\rm BR}(K^+\to\pi^+\nu\bar{\nu})$ are differently sensitive to
modifications from a given new-physics scenario.
If one or both BRs is found to differ from its SM value, it may be possible
to characterize the physical mechanism responsible, as schematized in
Figure~\ref{fig:bsm}, from \cite{Buras:2015yca}.
For example, if the pattern of flavor-symmetry breaking from new physics
were the same as in the SM (minimal flavor violation), the $K_L$ and $K^+$
BRs would lie along the band of correlation shown in green.
If the new interaction were to couple to only left-handed or only
right-handed quark currents, as expected for example in models with
modified $Z$ couplings or littlest Higgs models with $T$ parity, the
BRs would lie along one of the branches shown in blue.
New physics without these constraints, as expected for example
in models with large extra dimensions, could modify the $K^+$ and $K_L$ BRs
in an arbitrary way, as illustrated in red.

The decay ${\rm BR}(K_L\to\pi^0\nu\bar{\nu})$ has never been measured;
the KOTO experiment at J-PARC \cite{Ahn:2016kja} has a good chance 
of observing it. ${\rm BR}(K^+\to\pi^+\nu\bar{\nu})$ has been measured by 
Brookhaven experiment E787 and its successor, E949. The combined result
from the two generations of the experiment, obtained with seven candidate 
events, is ${\rm BR}(K^+\to\pi^+\nu\bar{\nu}) = 
1.73^{+1.15}_{-1.05}\times10^{-10}$ \cite{Artamonov:2009sz}. 
The purpose of the NA62 experiment at the CERN SPS is to measure 
${\rm BR}(K^+\to\pi^+\nu\bar{\nu})$ with a precision of about 10\%.
After cuts, the signal detection efficiency for such decays is on the
order of 10\%, so observation of $\sim$100 signal events will require
a sample of $10^{13}$ $K^+$ decays within the geometrical acceptance
of the experiment.
For a measurement with 10\% precision, the background
level must be kept down to no more than about 20\% of signal.
This implies an overall background rejection factor of $10^{12}$.

\section{The NA62 experiment}
\begin{figure}[ht]
\centering
\includegraphics[width=0.80\textwidth]{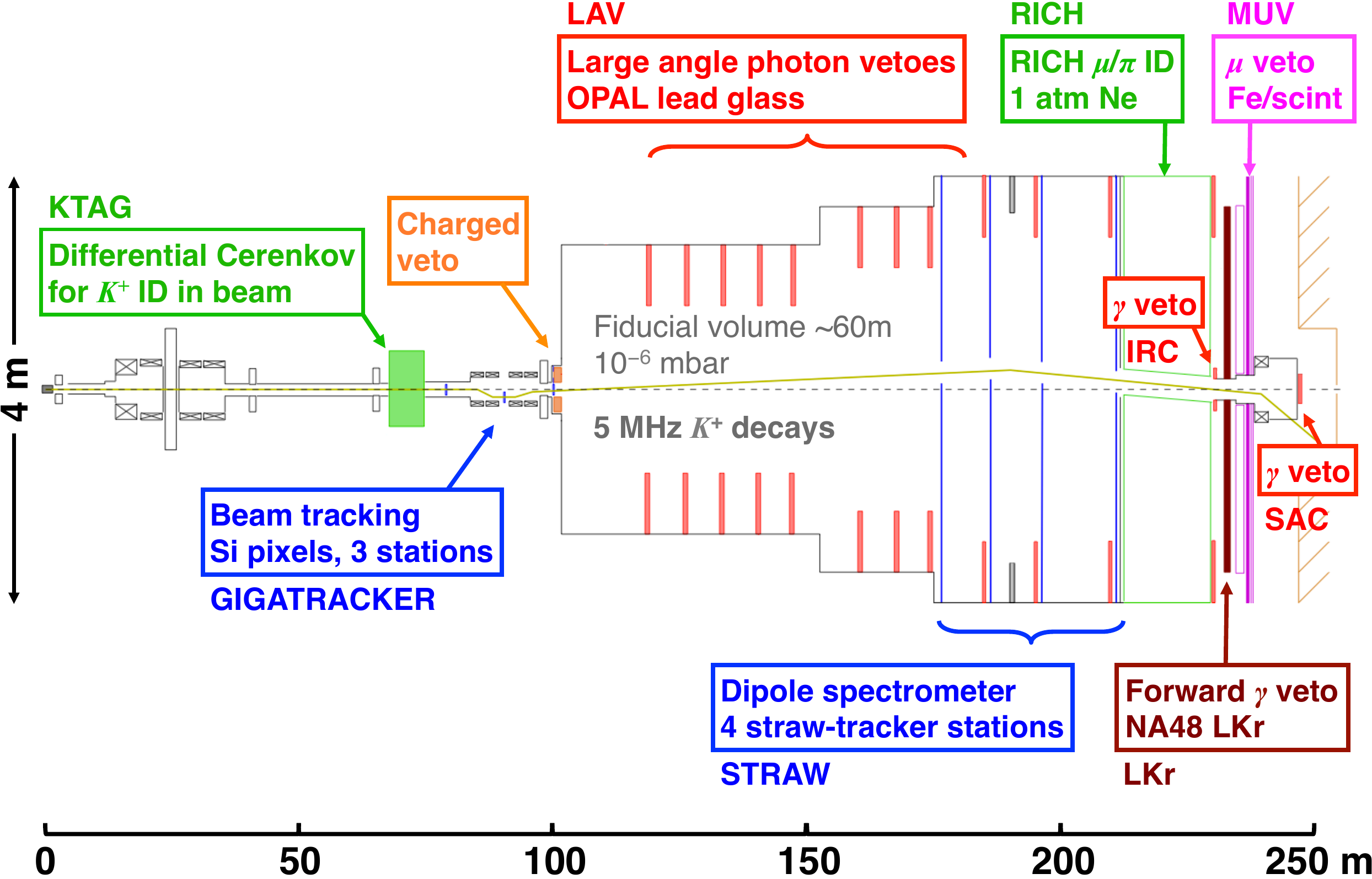}
\caption{Schematic diagram of the NA62 experiment.}
\label{fig:na62}
\end{figure}

The experimental signature for $K^+\to\pi^+\nu\bar{\nu}$ is a $K^+$
decaying to a $\pi^+$, with no other particles present.
The first line of defense against abundant decays such as
$K^+\to\mu^+\nu$ and $K^+\to\pi^+\pi^0$ (together representing about 84\% of
the total $K^+$ width) 
is to precisely reconstruct the missing mass of the primary and secondary
tracks and reject events with $m_{\rm miss}^2 \approx 0$ or
$m_{\rm miss}^2 \approx m_{\pi^0}^2$, assuming the secondary is
a $\mu^+$ or a $\pi^+$, respectively.
However, the rejection power from kinematics alone is not sufficient
(in NA62, it is at best $10^{4}$), and in any case, about 8\% of $K^+$ decays
(e.g., $K_{e3}$, $K_{\mu3}$) do not have 
closed kinematics. The remainder of the experiment's rejection power must 
come from redundant particle identification systems and hermetic, 
highly-efficient photon veto detectors. The NA62 apparatus \cite{NA62:2010xxx},
schematically illustrated in Figure~\ref{fig:na62}, was designed around 
these principles, which we now consider in turn.

\paragraph{Beamline and decay volume}
The experiment makes use of a 400-GeV primary proton beam from the SPS with 
$3\times10^{12}$ protons per pulse and a duty factor of about 0.3. The beam is 
collided on a beryllium target at zero angle to produce the 75-GeV $\pm1\%$ 
unseparated positive secondary beam used by the experiment. At production,
this beam consists of about 70\% $\pi^+$, 23\% $p$, and 7\% $K^+$, with a total
rate of 750~MHz. The beamline opens into the vacuum tank about 105~m
downstream of the target. The vacuum tank is about 115~m long and fully
encloses the four tracking stations of the magnetic spectrometer; the
pressure inside 
is kept at a level of $10^{-6}$ mbar. The fiducial volume occupies the
first 60~m of the vacuum tank, upstream of the spectrometer. About 
10\% of the $K^+$'s entering the experiment decay in the fiducial volume,
corresponding to 4.5~MHz of $K^+$ decays.

\paragraph{High-rate, precision tracking}
In order to obtain the full kinematic rejection factor of $10^{4}$ 
for two-body decays, both the beam particle and the decay secondary 
must be accurately tracked.
The beam spectrometer, referred to as the Gigatracker because it
tracks individual particles in the 750-MHz secondary beam,
consists of three hybrid silicon-pixel detectors
installed in an achromat in the beamline~\cite{Perrin-Terrin:2015lqj}.
The magnetic spectrometer for secondary particles consists of four straw
chambers operated inside the vacuum tank~\cite{Danielsson:2013qta}.
Each chamber has 16 layers of straw tubes arranged in 4 views.
A scintillator hodoscope downstream of the spectrometer allows
fast ($\sigma_t \sim 200$~ps) low-level triggering on charged particles.

\paragraph{Redundant particle identification}
The principal PID challenge for single tracks is to reject $K^+\to\mu^+\nu$ 
decays with an inefficiency of less than $10^{-7}$ after the application
of kinematic cuts. The bulk of NA62's $\pi/\mu$ separation capability is 
provided by the downstream muon vetoes (MUV). There are three MUV detectors. 
MUVs 1 and 2 are iron/scintillator hadron calorimeters. These are used 
mainly for offline $\pi/\mu$ separation and provide a muon rejection factor 
of $10^5$. MUV 3 is highly segmented and provides fast $\mu$
identification for triggering. An additional two orders of magnitude
of $\pi/\mu$
separation are provided by a large ($\sim$4-m-diameter by 17.5-m-long)
ring-imaging Cerenkov counter (RICH) \cite{Duk:2016bij} filled with
neon gas at 1~atm.

\paragraph{Beam timing and PID}
Considering that the rates of primary and secondary tracks in the experiment 
are respectively about 750~MHz and 10~MHz, accurately matching the correct
secondary track to the correct primary is not a trivial task.
Precise timing of the secondary can be obtained from the RICH
($\sigma_t \sim 100$~ps), while for the primary, the Gigatracker
provides $\sigma_t \sim 150$~ps.
The KTAG, a differential Cerenkov counter based on a CERN CEDAR-W detector
that has been refurbished and outfitted with a new, high-segmentation
readout~\cite{Goudzovski:2015xaa}, is used to identify kaons in the beam.
This both provides a precise ($\sigma_t \sim 100$~ps),
redundant measurement of the beam particle's timing and reduces the effective
beam rate from 750~MHz to 45~MHz, hence reducing the mismatch probability. 

\paragraph{Hermetic photon vetoes}
Rejection of photons from $\pi^0$'s is important for the elimination of 
many background channels, $K^+\to\pi^+\pi^0$ decays in particular.
For these decays, requiring the
$\pi^+$ to have $p < 35$~GeV guarantees that the two photons from the $\pi^0$
have a total energy of 40~GeV. If the missing-mass cuts provide a 
rejection power of $10^4$, the probability for the photon vetoes
to miss both photons must be less than $10^{-8}$.
The photon veto system consists of four separate subdetector systems.
The ring-shaped large-angle photon vetoes (LAVs) are placed at 12 locations 
along the vacuum volume and provide coverage for decay photons emitted
at angles to the beam line ($\theta$) between 8.5~and 50~mrad.
The LAV detectors consist of rings of lead-glass blocks salvaged from the 
OPAL electromagnetic calorimeter barrel \cite{Ambrosino:2011xxx}.
The NA48 liquid-krypton calorimeter (LKr),
a quasi-homogeneous ionization calorimeter of depth $27\,X_0$
and transverse segmentation of $2\times2$~cm$^2$ \cite{Fanti:2007vi},
vetoes forward ($1~{\rm mrad}<\theta<8.5~{\rm mrad}$), high-energy photons.
A ring-shaped shashlyk calorimeter (IRC) about the 
beamline provides coverage for photons with $\theta<1~{\rm mrad}$, while
further downstream, a small-angle shashlyk calorimeter (SAC) 
around which the beam is deflected completes the coverage for 
very-small-angle photons that would otherwise escape via the beam pipe.

\paragraph{Expected performance}
Simulations indicate that the acceptance for signal decays within the
fiducial volume is around 10\%, corresponding to about 45 signal events
accepted per year of data taking. The $\pi^+\pi^0$ background is
estimated to be about 10\% of signal counts, while the $\mu^+\nu$ background
is around 3\%. Including backgrounds from all other channels, the total
background is under 20\%.    

\section{Current status}

NA62 took its first data in a pilot run in the fall of 2014, and had
its first physics run during the summer and fall of 2015.
During the 2015 run, the beamline was commissioned up to the full design
intensity. The spectrometer, the PID detectors, and all of the photon veto
systems were fully operational; the Gigatracker was partially commissioned.
By the end of the 2015 run, the infrastructure for the FPGA-programmable
level-0 and software level-1 triggers was operational, with trigger conditions
and algorithms under study. A data set was collected at low beam intensity
(1\% of design) with a minimum bias trigger; this data was used to
study and verify the performance of the detector, as described below.
A data set collected at beam intensities of 50--100\% of the design
value was also collected and is currently under study.

\begin{figure}[ht]
\centering
\includegraphics[height=60mm]{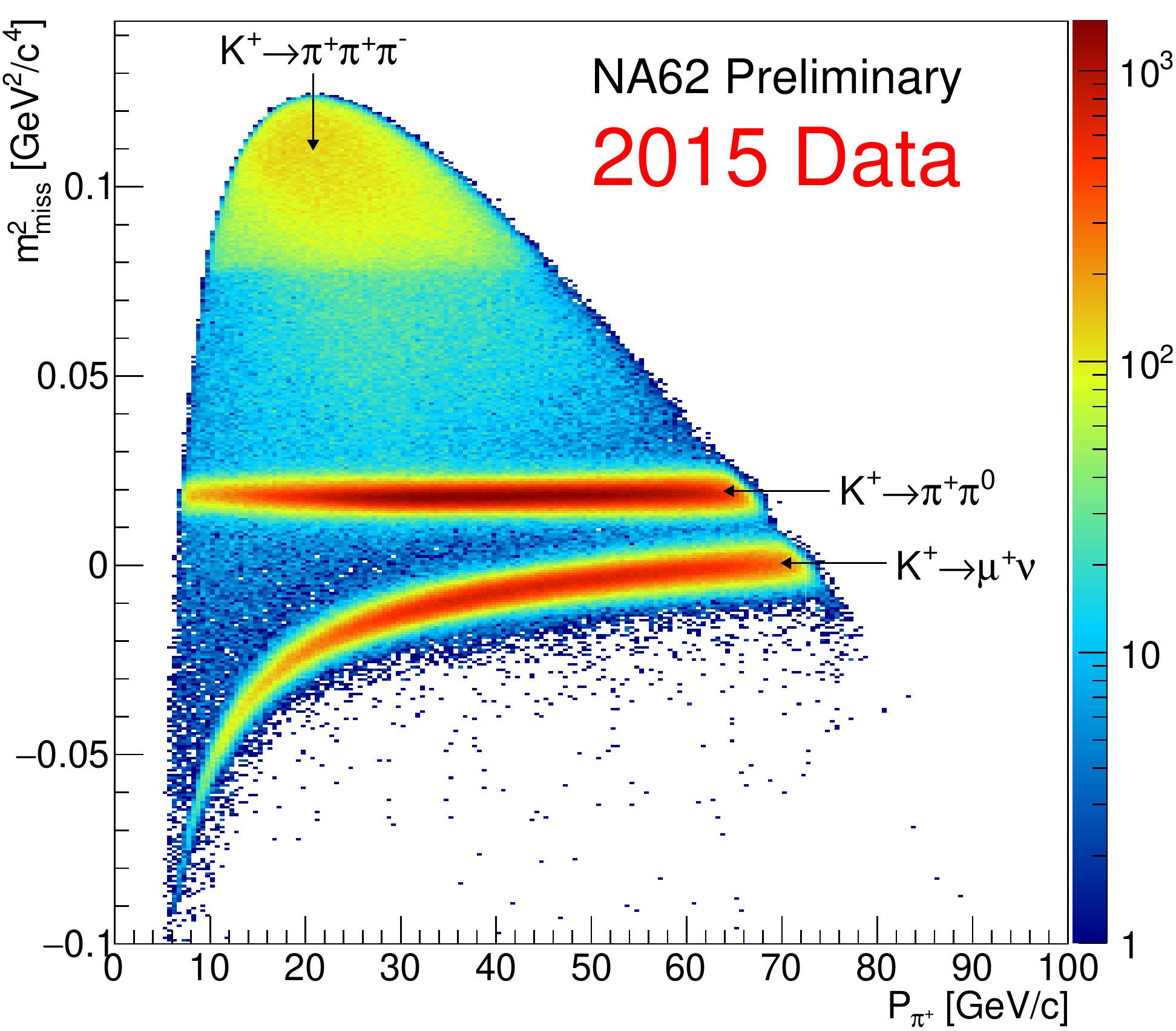}
\includegraphics[height=60mm]{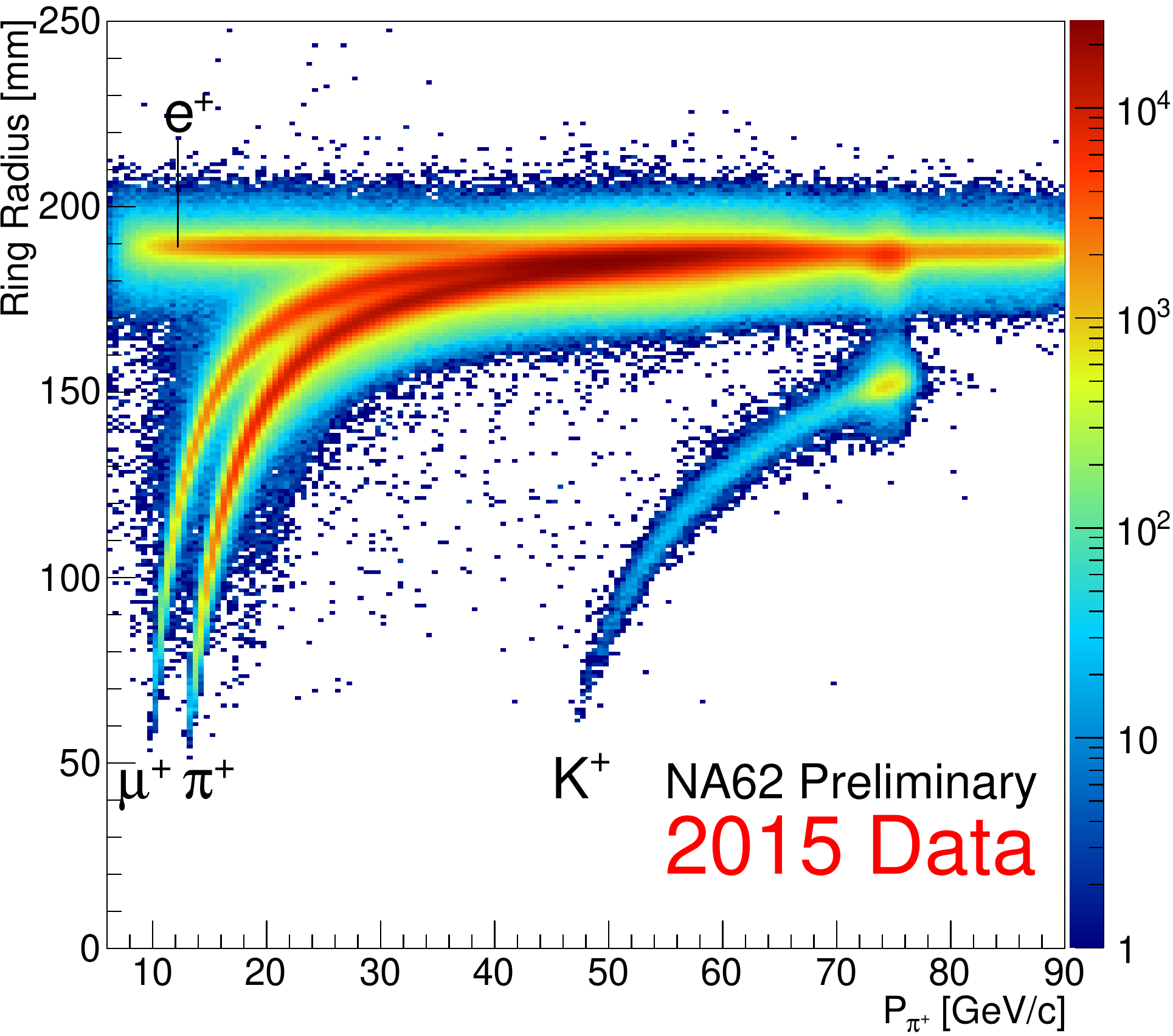}
\caption{Plots illustrating reconstruction quality for one-track events in
  NA62 data from the 2015 run. Left: squared missing mass $m^2_{\rm miss}$ at
  the presumed $K\pi$ vertex vs.\ track momentum $P_{\pi^+}$ measured in the
  spectrometer. Right: RICH ring radius vs.\ track momentum $P_{\pi^+}$.}
\label{fig:2015data}
\end{figure}
Selection of a sample of one-track events for efficiency and performance
studies proceeds as follows. Events with one positive track in the
spectrometer satisfying acceptance and quality cuts and with its origin
in the fiducial decay volume are first selected. This track is matched in
space and time to hits on the charged hodoscope, RICH rings, and hits
in the LKr and muon vetoes, if any. The track is also matched in time to any
hits on the photon vetoes or additional hits on the LKr. The spectrometer
track is then matched to a track from the Gigatracker, the vertex
is formed, and the track from the Gigatracker is matched in time with the
signal from the KTAG.

Figure~\ref{fig:2015data}, left, shows the distribution of squared missing mass
$m_{\rm miss}^2$ vs.\ track momentum $P_{\pi^+}$ for selected one-track events
with kaon beam identification from the KTAG.
$m_{\rm miss}^2$ is calculated assuming the track is from a $\pi^+$, and the
upper band contains $K^+\to\pi^+\pi^0$ events, which peak at
$m_{\rm miss}^2 = m_{\pi^0}^2$.
The lower band at $m_{\rm miss} \approx 0$ at high momentum and curving
towards negative values at low track momentum corresponds to the
$K^+\to\mu^+\nu$ events, for which the track mass hypothesis is incorrect.
The concentration of $K^+\to\pi^+\pi^+\pi^-$ events is visible at
$m_{\rm miss}^2 > 4m_{\pi^\pm}^2$. The region between the band from
$K^+\to\pi^+\pi^0$ and the concentration from $K^+\to\pi^+\pi^+\pi^-$
is populated by $K^+\to\pi^0e^+\nu$ and $K^+\to\pi^0\mu^+\nu$ decays.
Because the signal from the KTAG is required, few events are observed
outside the expected regions of correlation from abundant decays.
With the beam track reconstructed in the Gigatracker, for a clean
LKr-based selection of $K^+\to\pi^+\pi^0$
events, the missing-mass resolution is found to be
$\sigma(m_{\rm miss}^2) = 0.0013$ GeV$^2$, which is near to the design value,
and essentially independent of track momentum.

Figure~\ref{fig:2015data}, right, shows the distribution of ring radius
in the RICH vs.\ track momentum for selected tracks. The bands of correlation
for $e$, $\mu$, $\pi$, and $K$ are clearly visible. Pure samples of pions
and muons can be selected using kinematic cuts (see above) and allow
the $\pi/\mu$ separation power provided by the RICH and muon veto systems
to be measured. Preliminary results indicate that the RICH suppresses muons
by a factor of $\sim$10$^{-2}$ at 80\% efficiency for pions. From a separate
analysis, the suppression factor for muons provided by the muon vetoes is
$\sim$10$^{-6}$ at 50\% efficiency or better for pions; the pion efficiency
is expected to be further improved.
The tracking and particle identification systems can be used together to
obtain a clean sample of $K^+\to\pi^+\pi^0$ decays to study the photon
veto efficiency. The most directly relevant figure of merit is the overall
efficiency for detection of at least one of the two photons from the
$\pi^0$ in any of the photon vetoes. Thus defined, the $\pi^0$ rejection
is $\sim$10$^{-6}$, but this measurement is limited by the low statistics
of the 2015 minimum-bias data set.

\section{Broader physics program}

In addition to the measurement of ${\rm BR}(K^+\to\pi^+\nu\bar{\nu})$, 
the intense NA62 kaon beam and high-performance NA62 detector will allow
the pursuit of a broad program in kaon physics, including precision
measurements of dominant $K^+$ BRs, studies of decays of interest in
chiral perturbation theory, and a precision test of lepton universality
via the measurement of the ratio $R_K = \Gamma(K_{e2}/K_{\mu2})$.
The sample of $10^{13}$ $K^+$ decays will allow searches for lepton-flavor
(e.g. $K^+\to\pi^+\mu^\pm e^\mp$) or lepton-number
(e.g. $K^+\to\pi^-\ell^+\ell^+$)
violating decays with single-event sensitivities at the level of 10$^{-12}$.
Searches for heavy neutral leptons $\nu_{\rm h}$ can be performed either
inclusively,
via the missing-mass distribution for $K^+\to\ell^+\nu_{\rm h}$ decays
or via the direct search for $\nu_{\rm h}$ from upstream decays to
$\pi^\pm\ell^\mp$.
Long-lived dark sector particles produced in the target,
such as dark photons (decaying to $\ell^+\ell^-$) or axion-like particles
(decaying to $\gamma\gamma$) can also be searched for.
Finally, since ${\rm BR}(K^+\to\pi^+\pi^0) \approx 21\%$ and the daughter
$\pi^0$ is tagged by the reconstruction of the $K^+$ and $\pi^+$,
sensitive searches for rare or forbidden $\pi^0$ decays, such as
$\pi^0\to{\rm invisible}$, may also be carried out.

\section{Outlook}

The 2015 data show that, at low intensity, the detector is fully operational
and the key detector systems are largely performing as expected.
At the time of writing, the 2016 run is underway (April--November).
Additional runs are planned for 2017 and 2018. Work continues on optimizing
the trigger algorithms, readout performance, and reconstruction code for
running at higher intensities. Assuming the SM value for
${\rm BR}(K^+\to\pi^+\nu\bar{\nu})$, the experiment expects to collect a
few signal events by the end of 2016 run and is on track to collect $\sim$100
events and measure the BR to $\sim$10\% by 2018.


\begin{thebibliography}{99}

\bibitem{Cirigliano:2011ny} 
  V.~Cirigliano, G.~Ecker, H.~Neufeld, A.~Pich and J.~Portoles,
  Rev.\ Mod.\ Phys.\  {\bf 84}, 399 (2012)
  [arXiv:1107.6001 [hep-ph]].

\bibitem{Buras:2015qea} 
  A.~J.~Buras, D.~Buttazzo, J.~Girrbach-Noe and R.~Knegjens,
  JHEP {\bf 1511}, 033 (2015)
  [arXiv:1503.02693 [hep-ph]].

\bibitem{Buras:2015yca} 
  A.~J.~Buras, D.~Buttazzo and R.~Knegjens,
  JHEP {\bf 1511}, 166 (2015)
  [arXiv:1507.08672 [hep-ph]].

\bibitem{Ahn:2016kja}
  J.~K.~Ahn {\it et al.},
  arXiv:1609.03637 [hep-ex].

\bibitem{Artamonov:2009sz} 
  A.~V.~Artamonov {\it et al.}  [BNL-E949 Collaboration],
  Phys.\ Rev.\ D {\bf 79}, 092004 (2009)
  [arXiv:0903.0030 [hep-ex]].

\bibitem{NA62:2010xxx}
  F.~Hahn {\it et al.} (eds.)  [NA62 Collaboration],
  ``NA62: Technical Design Document,''
  NA62-10-07 (2010).
  
\bibitem{Perrin-Terrin:2015lqj} 
  M.~Perrin-Terrin [NA62 GTK Working Group],
  PoS VERTEX {\bf 2015}, 016 (2015).

\bibitem{Danielsson:2013qta} 
  H.~Danielsson [NA62 Collaboration],
  2013 IEEE Nuclear Science Symposium and Medical Imaging Conference,
  doi:10.1109/NSSMIC.2013.6829462.

\bibitem{Duk:2016bij} 
  V.~Duk,
  Int.\ J.\ Mod.\ Phys.\ Conf.\ Ser.\  {\bf 44}, 1660230 (2016).

\bibitem{Goudzovski:2015xaa} 
  E.~Goudzovski {\it et al.},
  Nucl.\ Instrum.\ Meth.\ A {\bf 801}, 86 (2015)
  [arXiv:1509.03773 [physics.ins-det]].
  
\bibitem{Ambrosino:2011xxx} 
  F.~Ambrosino {\it et al.},
  IEEE Nucl.\ Sci.\ Symp.\ Conf.\ Rec.\  {\bf 2011}, 1159 (2011)
  [arXiv:1111.4075 [physics.ins-det]].

\bibitem{Fanti:2007vi} 
  V.~Fanti {\it et al.}  [NA48 Collaboration],
  Nucl.\ Instrum.\ Meth.\ A {\bf 574}, 433 (2007).
 
\end{thebibliography}
\end{document}